%% file: proceedings_quantum_wave_analogies.tex
\documentclass[a4paper,10pt,numbers=noenddot,twoside=on,headinclude=true,footinclude=false,headsepline=true]{scrartcl}

\usepackage[utf8]{inputenc}
\usepackage[T1]{fontenc}
\usepackage{textcomp}
\usepackage[english]{babel}
\usepackage{color}
\usepackage{amssymb}
\usepackage{amsmath}
\usepackage{amsfonts}
\usepackage{amsopn}
\usepackage{amsbsy}
\usepackage{enumerate}
\usepackage[plainpages=false,pdfpagelabels,breaklinks=true,pdfborderstyle={/S/U/W 1.2}]{hyperref}
\usepackage{graphics}
\usepackage{dsfont}
\usepackage{units}
\usepackage[cal=boondoxo,bb=boondox]{mathalfa}

\numberwithin{equation}{section} 
\numberwithin{table}{section} 
\numberwithin{figure}{section} 

\usepackage[amsmath,thmmarks,thref,hyperref]{ntheorem}

\theoremstyle{plain}
\newtheorem{theorem}{Theorem}[section]
\newtheorem{definition}[theorem]{Definition}

\newtheorem{lemma}[theorem]{Lemma}

\newtheorem{proposition}[theorem]{Proposition}
\newtheorem{assumption}[theorem]{Assumption}

\theorembodyfont{\upshape}

\theoremstyle{nonumberplain}

\theoremsymbol{\ensuremath{_\Box}}
\newtheorem{proof}{Proof}

\usepackage[style=alphabetic,
	citestyle=alphabetic,
	firstinits=true,
	backend=biber,
	hyperref=auto,
	sorting=nyt,
	maxcitenames=3,
	maxbibnames=99,
	minnames=3,
	arxiv=pdf,
	url=false,
	isbn=false,
	dateabbrev=true,
	datezeros=true,
	bibencoding=utf8]{biblatex}
\newbibmacro{string+doi}[1]{%
	\iffieldundef{doi}{#1}{\href{http://dx.doi.org/\thefield{doi}}{#1}}}
\DeclareFieldFormat
	{title}{\usebibmacro{string+doi}{\mkbibemph{#1}}\isdot}
\DeclareFieldFormat
	[article,inbook,incollection,inproceedings,patent,thesis,unpublished]
	{title}{\usebibmacro{string+doi}{\mkbibemph{#1}}\isdot}
\DeclareFieldFormat
	[article,inbook,incollection,inproceedings,patent,thesis,unpublished]
	{pages}{#1}
\DeclareFieldFormat
	[article]
	{journaltitle}{#1\isdot}
\DeclareFieldFormat
	[article]
	{volume}{\textbf{#1}}
\DefineBibliographyStrings{english}{pages={}}
\DeclareBibliographyDriver{article}{%
	\usebibmacro{bibindex}%
	\usebibmacro{begentry}%
	\usebibmacro{author/translator+others}%
	\setunit{\labelnamepunct}\newblock
	\usebibmacro{title}%
	\usebibmacro{journal+issuetitle}%
	\setunit*{\addcomma\space}
	\printfield{pages}
	\setunit*{\addcomma\space}
	\printfield{year}
	\usebibmacro{finentry}%
}

\newbibmacro*{journal+issuetitle}{%
	\usebibmacro{journal}%
	\setunit*{\addspace}%
	\iffieldundef{series}
	{}
	{\newunit
		\printfield{series}%
		\setunit{\addspace}}%
	\iffieldundef{volume}
		{}
		{\printfield{volume}%
		\setunit{\addspace}}%
	\setunit*{\addcolon\space}%
	\usebibmacro{issue}%
	\newunit}

\usepackage[scaled]{beramono}
\usepackage{helvet}
\usepackage[charter]{mathdesign} 
\SetMathAlphabet{\mathcal}{normal}{OMS}{cmsy}{m}{n} 
\SetMathAlphabet{\mathcal}{bold}{OMS}{cmsy}{m}{n} 
\providecommand{\ie}{i.~e.~}
\providecommand{\eg}{e.~g.~}
\providecommand{\cf}{cf.~}

\providecommand{\R}{\mathbb{R}}

\providecommand{\C}{\mathbb{C}}
\renewcommand{\C}{\mathbb{C}}

\providecommand{\T}{\mathbb{T}}
\renewcommand{\T}{\mathbb{T}}
\providecommand{\N}{\mathbb{N}}
\providecommand{\Z}{\mathbb{Z}}
\providecommand{\ii}{\mathrm{i}}
\providecommand{\e}{\mathrm{e}}
\renewcommand{\Re}{\mathrm{Re} \,}
\renewcommand{\Im}{\mathrm{Im} \,}
\providecommand{\Hil}{\mathcal{H}}

\providecommand{\eps}{\varepsilon}
\providecommand{\Cont}{\mathcal{C}}

\providecommand{\ker}{\mathrm{ker} \, }
\providecommand{\ran}{\mathrm{ran} \, }

\providecommand{\ker}{\mathrm{ker} \,}

\providecommand{\dd}{\mathrm{d}}
\providecommand{\id}{\mathds{1}}

\providecommand{\order}{\mathcal{O}}

\providecommand{\norm}[1]{\left \lVert #1 \right \rVert}

\providecommand{\scpro}[2]{\left \langle #1 , #2 \right \rangle}

\providecommand{\bscpro}[2]{\bigl \langle #1 , #2 \bigr \rangle}

\providecommand{\sopro}[2]{\vert #1 \rangle \langle #2 \vert}

\providecommand{\Rot}{\mathrm{Rot}}
\providecommand{\Div}{\mathrm{Div}}

\usepackage{tabularx}

\bibliography{/Users/max/Library/texmf/tex/latex/max/bibliography}

\title{Taking Inspiration from \\ Quantum-Wave Analogies}
\subtitle{Recent Results for Photonic Crystals}
\author{Max Lein}

\begin{document}

\maketitle
\vspace{-9mm}
\begin{center}
	Advanced Institute of Materials Research,  
	Tohoku University \linebreak
	2-1-1 Katahira, Aoba-ku, 
	Sendai, 980-8577, 
	Japan \linebreak
	{\footnotesize \href{mailto:maximilian.lein.d2@tohoku.ac.jp}{\texttt{maximilian.lein.d2@tohoku.ac.jp}}}
\end{center}
\begin{abstract}
	Similarities between quantum systems and analogous systems for classical waves have been used to great effect in the physics community, be it to gain an intuition for quantum systems or to anticipate novel phenomena in classical waves. This proceeding reviews recent advances in putting these quantum-wave analogies on a mathematically rigorous foundation for classical electromagnetism. Not only has this \emph{Schrödinger formalism of electromagnetism} led to new, interesting mathematical problems for so-called Maxwell-type operators, it has also improved the understanding of the physics of topological phenomena in electromagnetic media. For example, it enabled us to classify electromagnetic media by their material symmetries, and explained why “fermionic time-reversal symmetries” — that were conjectured to exist in the physics literature — are in fact forbidden. 
\end{abstract}
%
%

\input{section_0}
\input{section_1}
\input{section_2}
\input{section_3}
\input{section_4}

\printbibliography

\end{document}

%% file: section_0.tex
\noindent
The idea behind quantum-wave analogies is that two effects, one in a quantum system and another in a classical wave, are in fact \emph{different manifestations of the same underlying physical principles.} They can be read both ways: The founding fathers of modern quantum mechanics relied on them to transfer some of their understanding of electromagnetism to their fledgling theory. Nowadays, it is usually the other way around, quantum mechanics is an established theory in its own right, and quantum-wave analogies can help guide one's intuition in the search for new phenomena in \emph{classical} waves. To name but one example, on the basis of the analogy to crystalline solids, Yablonovitch \cite{Yablonovitch:photonic_crystals_genisis:1987} and John \cite{John:photonic_crystals:1987} proposed the concept of \emph{photonic crystals}, man-made electromagnetic media whose periodic structure endows them with peculiar light conduction properties. This deceptively simple idea gave birth to several vibrant sub communities in physics. Despite very exciting experiments such as the realization of topological effects on a truly macroscopic scale \cite{Wang_et_al:unidirectional_backscattering_photonic_crystal:2009,Suesstrunk_Huber:mechanical_topological_insulator:2015} or the experimental observation of Weyl points \cite{Lu_et_al:experimental_observation_Weyl_points:2015}, only very few mathematicians are systematically working on classical wave equations with the help of quantum-wave analogies. That is a bit of a pity as new insights into the physics of classical waves open up very challenging mathematical questions which in some cases have a direct impact on the physics community. 

Least of which, not all of the quantum-wave analogies physicists have identified stand on a solid mathematical foundation, though, and it is not always clear whether and to what extent they can be made rigorous. Similarly, \emph{wave}-wave analogies are often not derived from first principles, because these different equations are usually not considered in a \emph{unified mathematical framework}. This proceeding will detail a mathematical framework that applies to a host of classical wave equations and discuss specific instances of quantum-wave analogies that were the subject of recent publications \cite{DeNittis_Lein:adiabatic_periodic_Maxwell_PsiDO:2013,DeNittis_Lein:symmetries_Maxwell:2014,DeNittis_Lein:ray_optics_photonic_crystals:2014,DeNittis_Lein:sapt_photonic_crystals:2013} and preprints \cite{DeNittis_Lein:Schroedinger_formalism_classical_waves:2017,DeNittis_Lein:symmetries_electromagnetism:2017}. Among other wave equations, the formalism applies to \emph{Maxwell's equations for linear dielectrics}, 
\begin{subequations}\label{intro:eqn:Maxwell_equations}
	\begin{align}
		\left (
		\begin{matrix}
			\eps & 0 \\
			0 & \mu \\
		\end{matrix}
		\right ) \, \frac{\partial}{\partial t} \left (
		\begin{matrix}
			\mathbf{E}(t) \\
			\mathbf{H}(t) \\
		\end{matrix}
		\right ) &= \left (
		\begin{matrix}
			+ \nabla \times \mathbf{H}(t) \\
			- \nabla \times \mathbf{E}(t) \\
		\end{matrix}
		\right ) - \left (
		\begin{matrix}
			\mathbf{J}(t) \\
			0 \\
		\end{matrix}
		\right )
		,
		\label{intro:eqn:Maxwell_equations:dynamics}
		\\
		\left (
		\begin{matrix}
			\nabla \cdot \eps \mathbf{E}(t) \\
			\nabla \cdot \mu \mathbf{H}(t) \\
		\end{matrix}
		\right ) &= \left (
		\begin{matrix}
			\rho(t) \\
			0 \\
		\end{matrix}
		\right )
		,
		\label{intro:eqn:Maxwell_equations:constraint}
		\\
		\partial_t \rho(t) + \nabla \cdot \mathbf{J}(t) &= 0 
		, 
		\label{intro:eqn:Maxwell_equations:charge_conservation}
	\end{align}
\end{subequations}
certain \emph{linearized magneto-hydrodynamic equations} \cite[Section~6.2.1]{DeNittis_Lein:Schroedinger_formalism_classical_waves:2017},
\begin{subequations}\label{intro:eqn:ideal_MHD}
	\begin{align}
		\frac{\dd \rho_1}{\dd t} &= - \nabla \cdot \bigl ( {\rho}_0 \, \textbf{v} \bigr )
		,
		\label{other_waves:eqn:ideal_MHD:1}
		\\
		\rho_0 \, \frac{\dd \textbf{v}}{\dd t} &= - \nabla \wp_1 - \textbf{g} \, \rho_1 + \bigl ( \textbf{J}_0 \times \textbf{B}_1 + \textbf{J}_1 \times \textbf{B}_0 \bigr )
		,
		\label{other_waves:eqn:ideal_MHD:2}
		\\
		\frac{\dd \textbf{B}_1}{\dd t} &= \nabla \times \bigl ( \textbf{v} \times \textbf{B}_0 \bigr )
		,
		\label{other_waves:eqn:ideal_MHD:3}
		\\
		\wp_1 &= \Gamma(\rho_1,\textbf{v})
		, 
		\label{other_waves:eqn:ideal_MHD:4}
	\end{align}
\end{subequations}
and linearized \emph{equations for spin propagating in a periodic medium} \cite[equation~(21)]{Shindou_et_al:chiral_magnonic_edge_modes:2013}, 
\begin{align}
	\ii \frac{\partial}{\partial t} \left (
	\begin{matrix}
		\beta(t,k) \\
		\overline{\beta(t,-k)} \\
	\end{matrix}
	\right ) = \sigma_3 \, H(k) \left (
	\begin{matrix}
		\beta(t,k) \\
		\overline{\beta(t,-k)} \\
	\end{matrix}
	\right ) 
	. 
	\label{intro:eqn:equations_spin_magnonic_crystal}
\end{align}
These equations share certain basic features: (1)~They are first-order in time, (2)~they have a product structure and (3)~physical waves are \emph{real}- as opposed to complex-valued. It turns out that all of these wave equations can be recast as a Schrödinger equation. 

%% file: section_1.tex
\section{The physics of photonic crystals} 
\label{sec:the_physics_of_photonic_crystals}
The big appeal of photonic crystals and periodic media for other classical waves is that they can be \emph{engineered} to have certain properties such as (photonic) band gaps and crystallographic symmetries. Here, physicists have much wider latitude than with condensed matter, not least because they are able to choose different spatial scales by varying the wavelength (\eg infrared or microwaves). 

The birth paper of the field of photonic crystals is the work by Yablonovitch \cite{Yablonovitch:photonic_crystals_genisis:1987} who proposed a dielectric with three-dimensional periodicity lattice and a \emph{photonic band gap} as an “omnidirectional quarter wave plate”. This inspired John \cite{John:photonic_crystals:1987} to suggest that also light can be Anderson localized in a photonic crystal with band gap in the presence of moderate disorder. 
These ideas were quickly generalized to other classical waves, and nowadays there are phononic crystals \cite{Fleury_et_al:breaking_TR_acoustic_waves:2014,Safavi-Naeini_et_al:2d_phonoic_photonic_band_gap_crystal_cavity:2014,Peano_Brendel_Schmidt_Marquardt:topological_phases_sound_light:2015,Chen_Zhao_Mei_Wu:acoustic_frequency_filter_topological_phononic_crystals:2017}, magnonic crystals \cite{Shindou_et_al:chiral_magnonic_edge_modes:2013} and periodic arrays of coupled oscillators \cite{Suesstrunk_Huber:mechanical_topological_insulator:2015,Suesstrunk_Huber:classification_mechanical_metamaterials:2016} whose physics is in many respects similar to that of crystalline solids.

\subsection{Photonic crystals with band gaps} 
\label{sub:photonic_crystals_with_band_gaps}
Initially, the focus of theoretical research centered around proposing dielectric structures which (1)~exhibited a photonic band gap and (2)~were thought to be realizable in experiment. Also robust numerical tools for band structure calculations have been developed \cite{Johnson_Joannopoulos:numerics_PhCs:2001}. 

Within the next decade or so, experimentalists pushed along two directions, namely improving the fidelity (absence of defects) and size (measured in numbers of unit cells, for instance) on the one hand, and the invention and improvement of manufacturing techniques for structures at smaller scales on the other (\cf \cite{Soukoulis:history_modelling_fabrication_PhC:2002,Kuramochi:fabrication_PhCs:2011} and \cite[Chapter~6]{Prather_et_al:PhCs:2009}). This means photonic crystals can now be designed and manufactured to have band gaps in specific, predetermined frequency ranges \cite{Joannopoulos_Johnson_Winn_Meade:photonic_crystals:2008}. 

\subsection{Topological photonic crystals} 
\label{sub:topological_photonic_crystals}
Late in 2005, Raghu and Haldane proposed in two seminal works \cite{Haldane_Raghu:directional_optical_waveguides:2008,Raghu_Haldane:quantum_Hall_effect_photonic_crystals:2008} to consider topological effects analogous to the Integer Quantum Hall Effect in adiabatically perturbed photonic crystals. Such slow modulations can be the result of strain \cite{Wong_et_al:strain_tunable_photonic_crystals:2004} or a thermal gradient \cite{Duendar_et_al:optothermal_tuning_photonic_crystals:2011,van_Driel_et_al:tunable_2d_photonic_crystals:2004}. Of particular interest are photonic crystals where the material weights $W = \left ( 
\begin{smallmatrix}
	\eps & 0 \\
	0 & \mu \\
\end{smallmatrix}
\right ) \neq \overline{W}$ are complex; this is analogous to magnetic quantum systems with broken time-reversal symmetry. Raghu and Haldane realized that the geometric (Berry) connection can be defined just like in crystalline solids and proposed a photonic analog of the \emph{Peierls substitution}: for states associated to an isolated band $\omega_n$ “semiclassical” ray optics equations of motion for $r$ and $k$ approximate the full light dynamics. Topological effects enter through the Berry curvature $\Omega_n := \nabla_k \times \ii \bscpro{\varphi_n}{\nabla_k \varphi_n}_w$. Based on this analogy, Haldane proposed the following \emph{photonic bulk-edge correspondence} \cite[p.~7]{Raghu_Haldane:quantum_Hall_effect_photonic_crystals:2008}: 
\begin{quote}
	[...] if the Chern number of a band changes at an interface, the net number of unidirectionally moving modes localized at the interface is given by the difference between the Chern numbers of the band at the interface (Fig.~2). 
\end{quote}
Initially, parts of the physics community were very skeptical that topological phenomena may present themselves in classical waves. Hence, it was only after the groups of Joannopoulos and Solja\v{c}i\'{c} at MIT provided conclusive experimental evidence \cite{Wang_et_al:edge_modes_photonic_crystal:2008,Wang_et_al:unidirectional_backscattering_photonic_crystal:2009} for the existence of unidirectional, backscattering-free edge modes, that Haldane was able to publish his two works. 

Haldane's insight has had a tremendous impact, it kickstarted the search for novel topological phenomena in photonic crystals \cite{Khanikaev_et_al:photonic_topological_insulators:2013} and other classical waves 
\cite{Fleury_et_al:breaking_TR_acoustic_waves:2014,Safavi-Naeini_et_al:2d_phonoic_photonic_band_gap_crystal_cavity:2014,Peano_Brendel_Schmidt_Marquardt:topological_phases_sound_light:2015,Chen_Zhao_Mei_Wu:acoustic_frequency_filter_topological_phononic_crystals:2017,Suesstrunk_Huber:mechanical_topological_insulator:2015,Suesstrunk_Huber:classification_mechanical_metamaterials:2016}. To give but one example, Lu et al carefully designed \cite{Lu_Fu_Joannopoulos_Soljacic:Weyl_points_gyroid_PhCs:2013} and realized \cite{Lu_et_al:experimental_observation_Weyl_points:2015} a photonic crystal with gyroid structures that exhibited Weyl points. Indeed, theirs was the first experimental observation of Weyl points (concurrently with a group at Princeton \cite{Xu_et_al:discovery_Weyl_points:2015} in a condensed matter system). For a comprehensive review of the physics we point the interested reader to \cite{Ozawa_et_al:review_topological_photonics:2018} or \cite{Khanikaev_Lein:review_topological_photonic_crystals:2018}. 

All of these table top experiments used microwaves. These are very convenient to work with for proof-of-principle experiments as their wavelength is in the millimeter to centimeter range: with yttrium iron garnet (YIG) there is a medium that is an \emph{insulator} and where the electric permittivity is \emph{complex} 
\cite{Pozar:microwave_engineering:2011}. Hence, we may regard the medium as being lossless for microwaves and $\eps \neq \overline{\eps}$ implies time-reversal symmetry is broken. Moreover, the degree to which it is broken can be tuned by applying a constant magnetic field. A second advantage is that these media can literally be machined \cite{Lu_et_al:experimental_observation_Weyl_points:2015} or, potentially, 3d printed 
\cite{Kadic_Bueckmann_Schittny_Wegener:metamaterials_beyond_electromagnetism:2013}. 

\subsection{Topological phenomena in media with time-reversal symmetry} 
\label{sub:topological_phenomena_in_media_with_time_reversal_symmetry}
However, for many applications is it desirable to use light at infrared or optical frequencies (\eg to realize “one-way wave guides” for photonic CMOS chips \cite{Gill_et_al:IBM_multiplexed_photonics_CMOS_chip:2015}). Fortunately, Maxwell's equations are scale-independent, and therefore the existence of topological phenomena for microwaves implies their existence for \emph{all other} frequencies — provided we are able to find media with the same material weights at those frequencies. 

This is where mathematics and physics part ways: currently, there are no known examples of  \emph{lossless} media with complex material weights at optical or infrared frequencies. Consequently, it is at present not feasible to miniaturize the experiments by Wang et al \cite{Wang_et_al:edge_modes_photonic_crystal:2008,Wang_et_al:unidirectional_backscattering_photonic_crystal:2009} or Lu et al \cite{Lu_et_al:experimental_observation_Weyl_points:2015}. Moreover, if one wants to deploy existing mass manufacturing techniques for processors, we would have to deal with time-reversal symmetric media like silicon or gallium arsenide. 

Wu and Hu \cite{Wu_Hu:topological_photonic_crystal_with_time_reversal_symmetry:2015} proposed a way to realize topological effects in ordinary, time-reversal symmetric media with the help of lattice symmetries. The idea is that the presence of the extra lattice symmetry gives rise to an “isospin” degree of freedom, and we may consider isospin-$\uparrow$ and isospin-$\downarrow$ systems separately. And as time-reversal symmetry flips the isospin, time-reversal symmetry is broken \emph{on each subsystem}. Thus, if edge modes exist, they come in counter-propagating pairs of opposite isospin. When states of a given isospin are selectively excited, then they feature the same hallmarks of topological protection as those observed in \cite{Wang_et_al:unidirectional_backscattering_photonic_crystal:2009}. However, these unidirectional edge modes are only robust against perturbations which \emph{preserve the relevant crystallographic symmetries}. Their idea has since been realized in a variety of experiments \cite{Wu_et_al:experiment_topological_photonic_crystal_with_time_reversal_symmetry:2017,Ye_et_al:experiment_topological_photonic_crystal_with_time_reversal_symmetry:2017}. The significance here is that time-reversal symmetric media are plentiful at many wave lengths, and exploiting crystallographic symmetries provides a viable avenue to engineer media with topological phenomena based on readily available materials. 

\subsection{Periodic waveguide arrays} 
\label{physics:periodic_waveguide_arrays}
Another class of experiments with optical and infrared light is performed with periodic waveguide arrays, where topologically protected edge states have also been observed \cite{Garanovich_et_al:modulated_photonic_lattices:2012,Plotnik_et_al:edge_modes_photonic_graphene:2012,Plotnik_Rechtsman_et_al:unconventional_edge_states_photonic_graphene:2014,Guzman-Silva_et_al:experiment_bulk_edge_photonic_Lieb_lattice:2014}. Here, the waveguides are formed by tubular regions where the refractive index $n$ of the substrate material (usually silica where $n \approx 1.45$) is increased by $\Delta n \approx 10^{-3} \sim 10^{-4}$. For light of specific wavelengths, these regions of increased refractive index act as \emph{evanescently coupled waveguides}. By carefully designing these waveguide arrays, experimentalists can, to good approximation, realize a variety of tight-binding Hamiltonians with nearest and next-nearest neighbor hopping, \eg “photonic graphene” \cite{Plotnik_Rechtsman_et_al:unconventional_edge_states_photonic_graphene:2014} or a Lieb lattice \cite{Guzman-Silva_et_al:experiment_bulk_edge_photonic_Lieb_lattice:2014}. The wave guide array geometry can be modified in other ways, say with helically wound waveguides and “squeezed” configurations \cite{Rechtsman_Plotnik_et_al:edge_states_photonic_graphene:2013}, and its effect on light conduction properties have been investigated. Physicists propose the paraxial wave equation 
\begin{align}
	\ii \partial_z \psi(x,z) &= - \frac{1}{2 k_0} \Delta_x \psi(x,z) - \frac{k_0 \, \Delta n(x)}{n_0} \psi(x,z)
	\label{physics:eqn:paraxial_Schroedinger_eqn}
\end{align}
as the equation which governs the “dynamics”. Here, $z$ is the coordinate along the fiber direction which takes the role of time, $x = (x_1,x_2)$ are the coordinates orthogonal to the $z$-axis, and $\psi$ is the envelope of the electromagnetic field, \ie $\mathbf{E}(t,x,z) = \Re \psi(x,z) \; \e^{+ \ii (k_0 z - \omega t)}$. The operator which appears on the right-hand side of \eqref{physics:eqn:paraxial_Schroedinger_eqn} has the same form as a regular Schrödinger operator; sometimes a pseudomagnetic field and other potential terms are added to phenomenologically account for geometric effects \cite{Longhi:quantum_optical_ananlog_review:2009,Rechtsman_Zeuner_et_al:photonic_topological_insulators:2013}. These waveguide arrays can be endowed with “artificial dimensions” by making them parameter-dependent; in recent experiments \cite{Ozawa_et_al:synthetic_dimensions_photonics_4d_quantum_Hall_effect:2016,Zilberberg_et_al:4d_quantum_Hall_effect_waveguide_array:2017} topological phenomena have been found in two-dimensional boundaries to effectively four-dimensional waveguide arrays. 

Seeing as the effective equation~\eqref{physics:eqn:paraxial_Schroedinger_eqn} describing waveguide arrays is different from Maxwell's equations, we will not cover them any further in these proceedings. Instead, we refer to \cite[Section~III.A.2]{Ozawa_et_al:review_topological_photonics:2018} for details. 

%% file: section_2.tex
\section{The Schrödinger formalism for classical waves} 
\label{Schroedinger}
When one wants to make a specific quantum-wave analogy rigorous, rewriting wave equations such as \eqref{intro:eqn:Maxwell_equations}–\eqref{intro:eqn:equations_spin_magnonic_crystal} in the form 
\begin{align}
	\ii \partial_t \Psi(t) = M \Psi(t) - \ii J(t) 
	, 
	&&
	\Psi(t_0) = \Phi \in \Hil_W
	,
	\label{Schroedinger:eqn:Schroedinger_equation}
\end{align}
is a natural preliminary step that necessarily precedes anything else. Here, $\Psi(t)$ is a \emph{complex} wave that \emph{represents} the \emph{real}-valued physical wave $u(t) = 2 \Re \Psi(t)$ and belongs to a \emph{complex} Hilbert space $\Hil_W$; the selfadjoint operator $M$ takes the place of the quantum Hamiltonian; and $J(t)$ is a current. The derivation of all the pieces in equation~\eqref{Schroedinger:eqn:Schroedinger_equation} consists of two distinct steps:
\begin{enumerate}[(1)]
	\item We complexify the wave equation and obtain an equation that looks like \eqref{Schroedinger:eqn:Schroedinger_equation}; this gives us access to many of the powerful tools developed for selfadjoint operators. 
	\item We reduce the complexified equations to waves with frequencies $\omega \geq 0$. Then real-valued physical waves $u = 2 \Re \Psi$ are \emph{uniquely represented} by a complex wave $\Psi$ composed solely of non-negative frequencies. 
\end{enumerate}

\subsection{Maxwell-type operators for the complexified equations} 
\label{Schroedinger:complexification}
What makes these \emph{Maxwell-type operators} $M^{\C} = W_L \, D \, W_R$, which arise in the context of wave equations, mathematically interesting is their \emph{product structure.}
\begin{definition}[Complexified Maxwell-type operator]\label{other_waves:defn:Maxwell_type_operator}
	A linear operator
	\begin{align}
		M^{\C} := W_L \, D \, W_R
		\label{Schroedinger:eqn:generalized_acoustic_operator}
	\end{align}
	with product structure on a complex Hilbert space $\Hil^{\C}$ is a complexified Maxwell-type operator if it has the following properties: 
	\begin{enumerate}[(a)]
		\item $D$ is a (possibly unbounded) selfadjoint operator on $\Hil^{\C}$ with domain $\mathcal{D}_0$. 
		\item $W_L , W_R \in \mathcal{B}(\Hil^{\C})$ are bounded, selfadjoint, commuting operators with bounded inverses, \ie $[W_L , W_R] = 0$ and $W_L^{-1} , W_R^{-1} \in \mathcal{B}(\Hil^{\C})$. 
		\item The product $W_R \, W_L^{-1}$ is selfadjoint and bounded away from $0$, \ie there exists a positive constant $c > 0$ so that $W_R \, W_L^{-1} \geq c \, \id_{\Hil^{\C}}$. 
		\item $M^{\C}$ is endowed with the domain $\mathcal{D}_R := W_R^{-1} \, \mathcal{D}_0$. 
		\item $M^{\C}$ anticommutes with an even antiunitary conjugation $C$ (which we will refer to as \emph{complex conjugation}), \ie $C \, \mathcal{D}_R = \mathcal{D}_R$ is left invariant and $M^{\C}$ satisfies 
		\begin{align}
			C \, M^{\C} \, C = - M^{\C}
			. 
			\label{Schroedinger:eqn:particle_hole_symmetry_M_C}
		\end{align}
	\end{enumerate}
\end{definition}
Straightforward arguments show that Maxwell-type operators are necessarily closed \cite[Lemma~B.1]{DeNittis_Lein:Schroedinger_formalism_classical_waves:2017}; however, they generally fail to be symmetric on $\Hil^{\C}$ and therefore cannot be selfadjoint. This apparent lack of selfadjointness can be cured by endowing the \emph{vector} space $\Hil^{\C}$ with the \emph{weighted} scalar product 
\begin{align}
	\bscpro{\Phi}{\Psi}_W := \bscpro{\Phi}{W_L^{-1} \, W_R \, \Psi}
	\label{Schroedinger:eqn:weighted_scalar_product}
\end{align}
defined in terms of $\Hil^{\C}$'s scalar product $\scpro{\, \cdot \,}{\, \cdot \,}$. We will denote the resulting Hilbert space with $\Hil_W^{\C}$; due to our assumptions on $W_L$ and $W_R$, the vector spaces $\Hil^{\C}_W$ and $\Hil^{\C}$ agree as \emph{Banach} spaces \cite[Section~6.1.1]{DeNittis_Lein:Schroedinger_formalism_classical_waves:2017}. A quick computation shows that Maxwell-type operators are indeed symmetric on $\Hil_W$, 
\begin{align*}
	\bscpro{\psi}{M \varphi}_W &= \bscpro{\psi}{W_R \, D \, W_R \varphi}
	= \bscpro{W_R \, D \, W_R \psi}{\varphi}
	\\
	&= \bscpro{W_L^{-1} \, W_L \, D \, W_R \psi}{W_R \varphi}
	= \bscpro{M \psi}{\varphi}_W 
	, 
\end{align*}
and elementary arguments in fact lead us to conclude that $M$ coincides with its $\scpro{\, \cdot \,}{\, \cdot \,}_W$-adjoint $M^{\ast_W}$ \cite[Proposition~6.2]{DeNittis_Lein:Schroedinger_formalism_classical_waves:2017}. Thanks to that, we can employ functional calculus \eg to define the \emph{unitary} evolution group $\e^{- \ii t M^{\C}}$. Note that the unitarity leads to the existence of at least one conserved quantity, $\mathcal{E}(\psi) = \tfrac{1}{2} \scpro{\psi}{\psi}_W$. 

The role of the symmetry relation~\eqref{Schroedinger:eqn:particle_hole_symmetry_M_C} is to ensure the existence of real solutions: $C \, M^{\C} \, C = - M^{\C}$ implies that complex conjugation $C$ and therefore, the associated real part operator $\Re = \tfrac{1}{2} (\id + C)$, commute with the unitary evolution group, 
\begin{align*}
	\Re \, \e^{- \ii t M^{\C}} = \e^{- \ii t M^{\C}} \, \Re 
	. 
\end{align*}
Put succinctly, the fact that the original equations describe real waves translates to the presence of the symmetry $C$. Moreover, it gives us a way to represent real waves as the real part of complex waves: if $\Psi \in \Hil^{\C}_W$ is a complex wave with $u = 2 \Re \Psi$, then we may either evolve $u$ or evolve the complex wave $\Psi$ and then take $2 \Re$ afterwards. 

\subsection{Reduction to complex waves of non-negative frequencies} 
\label{Schroedinger:reduction_complex_waves}
This insight is crucial for the second step where we establish a one-to-one correspondence between real, physical waves and complex waves composed only of frequencies with $\omega \geq 0$. 
In principle, there are many different one-to-one mappings $u \mapsto \Psi$ with $u = 2 \Re \Psi$, but spectral conditions are particularly easy to impose and so an especially convenient choice. For the \emph{in vacuo} Maxwell equations this strategy is part and parcel of every course on electromagnetism, but it is worth generalizing them to cases where solutions to Maxwell's equations can no longer be explicitly expanded in terms of pseudo eigenfunctions. 

Any real field $u = \Re u \in \Hil_W^{\C}$ can be decomposed via the maps 
\begin{align}
	Q_{\pm} = 1_{(0,\infty)}(\pm M^{\C}) + \tfrac{1}{2} \, 1_{\{ 0 \}}(M^{C}) 
	= \id - Q_{\mp} 
	= C \, Q_{\mp} \, C 
	\label{Schroedinger:eqn:Q_pm}
\end{align}
into a non-negative and a non-positive frequency component, 
\begin{align*}
	u = \Psi_+ + \Psi_- 
	:= Q_+ u + Q_- u 
	. 
\end{align*}
Thus, for real waves $u$ the non-positive frequency part $\Psi_- = Q_- u$ can be reconstructed from the non-negative frequency contribution 
\begin{align}
	\Psi_- = Q_- u = \overline{\Psi_+} 
	= C \, Q_+ u
	\label{Schroedinger:eqn:phase_locking_condition}
\end{align}
by taking the complex conjugate. This \emph{phase locking condition} tells us that $\Psi_+$ and $\Psi_-$ are \emph{not independent degrees of freedom}, and that the real-valued physical fields 
\begin{align*}
	u = 2 \Re \Psi_{\pm}
\end{align*}
can be recovered from just $\Psi_{\pm} \in \ran Q_{\pm}$ alone. Conversely, $u$ uniquely determines $\Psi_{\pm} = Q_{\pm} u$, and a straight-forward adaption of the arguments that led to \cite[Proposition~3.3]{DeNittis_Lein:Schroedinger_formalism_classical_waves:2017} yields 
\begin{proposition}[1-to-1 correspondence between real and complex $\omega \geq 0$ waves]\label{Schroedinger:prop:1_to_1_correspondence}~\\
	Given a complexified Maxwell operator $M^{\C} = W_L \, D \, W_R$, let $Q_{\pm}$ denote the maps from equation~\eqref{Schroedinger:eqn:Q_pm}. Then for real states $2 \Re$ is an inverse of $Q_{\pm}$ in the sense that 
	\begin{align*}
		2 \Re \, Q_{\pm} \, \Re = \Re : \Hil_W^{\C} \longrightarrow \Hil_W^{\C} 
	\end{align*}
	holds. Therefore, there is a one-to-one correspondence between the real wave $u = \Re u = 2 \Re \Psi_{\pm} \in \Hil_W^{\C}$ and the complex wave $\Psi_{\pm} = Q_{\pm} u \in \Hil_W^{\C}$. 
\end{proposition}
As a matter of convention, real waves are represented as complex waves comprised of non-negative frequencies. Due to this correspondence, we can describe the real wave equation for $u = 2 \Re \Psi_+$ as a Schrödinger-type equation for the complex wave $\Psi_+$, 
\begin{align}
	\ii \partial_t \Psi_+(t) = M_+ \Psi_+(t) - \ii J_+(t) 
	, 
	&&
	\Psi_+(t_0) = Q_+ u(t_0) \in \Hil_W
	, 
	\label{Schroedinger:eqn:Schroedinger_equation_positive_frequencies}
\end{align}
where the Hilbert space $\Hil_W = Q_+ \bigl [ \Hil_W^{\C} \bigr ]$ is the \emph{non-negative frequency subspace of the Hilbert space $\Hil_W^{\C}$} that inherits the weighted scalar product~\eqref{Schroedinger:eqn:weighted_scalar_product}, and the Maxwell-type operator 
\begin{align}
	M_+ = W_L \, D \, W_R \, \big \vert_{\omega \geq 0} 
	:= M^{\C} \, \big \vert_{\Hil_W}
	\label{Schroedinger:eqn:Maxwell_operator}
\end{align}
is the restriction of the complexified Maxwell operator to the spectral subspace $\omega \geq 0$. Note that all essential properties transfer from $M^{\C}$, \eg as a restriction of a selfadjoint operator to a spectral subspace, it is automatically selfadjoint. 

The physical, real-valued wave 
\begin{align*}
	u(t) = 2 \Re \Psi_+(t) 
\end{align*}
is recovered by taking twice the real part. 

Note that the selfadjointess of $M_+$ leads to the existence of at least one conserved quantity in the absence of sources, the square of the weighted norm, 
\begin{align*}
	\mathcal{E} \bigl ( u(t) \bigr ) := \scpro{\e^{- \ii t M_+} Q_+ u \, }{ \, \e^{- \ii t M_+} Q_+ u}_W
	= \bscpro{Q_+ u \, }{ \, Q_+ u}_W
	= \mathcal{E} \bigl ( u(0) \bigr )
	\, . 
\end{align*}
In case of Maxwell's equations, this conserved quantity is nothing but the total field energy. 

\subsection{Changes of representation} 
\label{Schroedinger:change_of_representation}
One of the concepts that directly transfers from quantum mechanics is that of \emph{representations}; this can be useful to exploit extra structures, symmetries or other relations. Unitary operators between $\Hil_W$ and other Hilbert spaces facilitate changes of representation. The two most common examples are the continuous Fourier transform to analyze homogeneous media or a version of the discrete Fourier transform for periodic media. 

Another relevant example is $W_R : \Hil_W \longrightarrow \Hil_{W'}$, viewed as a \emph{unitary} operator between $\Hil_W$ and the Hilbert space $\Hil_{W'} = W_R [\Hil_W]$, where the latter is endowed with the scalar product weighted by the operator $W' := W_L^{-1} \, W_R^{-1}$, 
\begin{align*}
	\bscpro{\Phi}{\Psi}_{W'} = \bscpro{\Phi \, }{ \, W_L^{-1} \, W_R^{-1} \, \Psi}
	. 
\end{align*}
The advantage here is that $W_R$ transforms a Maxwell-type operator of the form~\eqref{Schroedinger:eqn:generalized_acoustic_operator} to one where the weights are all on the left, 
\begin{align*}
	M = {W'}^{-1} \, D 
	\, . 
\end{align*}
Consequently, we may take this to be the canonical form of Maxwell-type operators. 

\subsection{The Schrödinger formalism for Maxwell's equations for linear, dispersionless media} 
\label{Schroedinger:Maxwell_equations}
Let us now make the construction explicit for the case of Maxwell's equations that govern the propagation of electromagnetic fields $\bigl ( \mathbf{E}(t) , \mathbf{H}(t) \bigr ) \in L^2(\R^3,\R^6)$ in a non-gyrotropic dielectric. Here, the material weights are constructed from the electric permittivity $\eps$ and the magnetic permeability $\mu$, 
\begin{align*}
	W := \left (
	\begin{matrix}
		\eps & 0 \\
		0 & \mu \\
	\end{matrix}
	\right )
	. 
\end{align*}
Non-gyrotropic dielectrics are those where the weights $W = \overline{W}$ are real and satisfy 
\begin{assumption}[Material weights]\label{Schroedinger:assumption:material_weights}
	The medium described by the material weights $W \in L^{\infty} \bigl ( \R^3 , \mathrm{Mat}_{\C}(6) \bigr )$ has the following properties: 
	\begin{enumerate}[(a)]
		\item The medium is \emph{lossless}, \ie $W(x) = W(x)^*$ takes values in the \emph{hermitian} matrices. 
		\item The medium is \emph{not a negative index material}, \ie for some $C > c > 0$ the weights satisfy 
		\begin{align*}
			0 < c \, \id \leq W \leq C \, \id < \infty 
			. 
		\end{align*}
	\end{enumerate}
\end{assumption}

\subsubsection{Complexified equations} 
\label{Schroedinger:Maxwell_equations:complexified}
Under these conditions, we can multiply both sides of the dynamical Maxwell equation~\eqref{intro:eqn:Maxwell_equations:dynamics} with $\ii \, W^{-1}$ and obtain the Schrödinger-type equation 
\begin{align}
	\ii \, \partial_t \Psi(t) = M^{\C} \, \Psi(t) - \ii \, W^{-1} \, \bigl ( \mathbf{J}(t) , 0 \bigr )
	\label{Schroedinger:eqn:Schroedinger_Maxwell_complexified}
\end{align}
with the initial condition 
\begin{align*}
	\Psi(t_0) = \bigl ( \mathbf{E}(t_0) , \mathbf{H}(t_0) \bigr ) \in L^2(\R^3,\R^6) \subset L^2(\R^3,\C^6)
	. 
\end{align*}
Here, $\Psi(t) = \bigl ( \mathbf{E}(t) , \mathbf{H}(t) \bigr )$ is just a new label for the electromagnetic field and the complexified Maxwell operator 
\begin{align*}
	M^{\C} = W^{-1} \, \Rot
	:= \left (
	\begin{matrix}
		\eps & 0 \\
		0 & \mu \\
	\end{matrix}
	\right )^{-1} \, \left (
	\begin{matrix}
		0 & + \ii \nabla^{\times} \\
		- \ii \nabla^{\times} & 0 \\
	\end{matrix}
	\right )
\end{align*}
is a succinct way to write the right-hand side of \eqref{Schroedinger:eqn:Schroedinger_Maxwell_complexified}. The \emph{free Maxwell operator} $\Rot$ is defined in terms of the curl $\nabla^{\times} \mathbf{E} := \nabla \times \mathbf{E}$. On the usual $L^2(\R^3,\C^6)$ the complexified Maxwell operator is closed, but not selfadjoint; however, if we endow this Banach space with the (weighted) \emph{energy scalar product} 
\begin{align*}
	\scpro{\Phi}{\Psi}_W := \int_{\R^3} \dd x \, \Phi(x) \cdot W(x) \Psi(x) 
	, 
\end{align*}
and denote the resulting Hilbert space with $\Hil_W^{\C} := L^2_W(\R^3,\C^6)$, then $M^{\C}$ endowed with the domain of $\Rot$ indeed defines a selfadjoint operator on $\Hil_W^{\C}$ \cite[Proposition~6.2]{DeNittis_Lein:Schroedinger_formalism_classical_waves:2017}; note that complex conjugation in the above expression is contained in the Euclidean scalar product $a \cdot b := \sum_{j = 1}^6 \overline{a_j} \, b_j$ of $\C^6$. 

This takes care of the dynamical equation, but we must not forget about the \emph{constraint equation~\eqref{intro:eqn:Maxwell_equations:constraint}}. To see that this is also satisfied, we introduce a Helmholtz splitting 
\begin{align*}
	\Hil^{\C} = \mathcal{J} \oplus \mathcal{G} 
\end{align*}
into transversal and longitudinal waves that is adapted to the medium. Here, longitudinal waves are those which are static, \ie gradient fields 
\begin{align*}
	\mathcal{G} := \Bigl \{ \bigl ( \nabla \varphi^E , \nabla \varphi^H \bigr ) \in L^2(\R^3,\C^6) \; \; \big \vert \; \; \varphi^E , \varphi^H \in L^2_{\mathrm{loc}}(\R^3) \Bigr \} 
	= \ker \Rot 
	= \ker M^{\C}
	, 
\end{align*}
whereas transversal states are those which are $\scpro{\, \cdot \,}{\, \cdot \,}_W$-orthogonal to them, 
\begin{align*}
	\mathcal{J} := \mathcal{G}^{\perp_W} 
	= \bigl \{ \Psi \in L^2(\R^3,\C^6) \; \; \big \vert \; \; \Div \, W \Psi = 0 \bigr \} 
	= \ran M^{\C} 
	, 
\end{align*}
where $\Div (\mathbf{E},\mathbf{H}) := \bigl ( \nabla \cdot \mathbf{E} \, , \, \nabla \cdot \mathbf{H} \bigr )$ consists of two copies of the usual divergence operator. Transversal states $\Psi_{\perp} \in \mathcal{J}$ are exactly those that satisfy the constraint equation~\eqref{intro:eqn:Maxwell_equations:constraint} in the distributional sense for $\rho = 0$. 

Applying this decomposition to the Schrödinger-type equation yields that the longitudinal part $\Psi_{\parallel} \in \mathcal{G}$ is fixed by the constraint equation~\eqref{intro:eqn:Maxwell_equations:constraint} (\cf \cite[Section~3.2.1]{DeNittis_Lein:Schroedinger_formalism_classical_waves:2017}), and only the transversal part $\Psi_{\perp} \in \mathcal{J}$ has non-trivial dynamics. Consequently, given sources $\rho(t)$ and $\mathbf{J}(t)$ which satisfy local charge conservation~\eqref{intro:eqn:Maxwell_equations:charge_conservation} and an initial condition $\bigl ( \mathbf{E}(t_0) , \mathbf{H}(t_0) \bigr )$ that fulfills the constraint equation~\eqref{intro:eqn:Maxwell_equations:constraint}, then also the solution 
\begin{align*}
	\bigl ( \mathbf{E}(t) , \mathbf{H}(t) \bigr ) &= \e^{- \ii (t - t_0) M^{\C}} \bigl ( \mathbf{E}(t_0) , \mathbf{H}(t_0) \bigr ) - \ii \int_{t_0}^t \dd s \, \e^{- \ii (t - s) M^{\C}} \, W^{-1} \, \bigl ( \mathbf{J}(t) , 0 \bigr ) 
\end{align*}
to the Schrödinger equation~\eqref{Schroedinger:eqn:Schroedinger_Maxwell_complexified} satisfies \eqref{intro:eqn:Maxwell_equations:constraint} for all times. 

\subsubsection{Reduction to $\omega \geq 0$ and one-to-one correspondence} 
\label{Schroedinger:Maxwell_equations:reduction}
Because this operator is selfadjoint, the operator $Q_+$ defined through \eqref{Schroedinger:eqn:Q_pm} makes sense and can be used to define the non-negative frequency subspace 
\begin{align*}
	\Hil_W := Q_+ \bigl [ L^2_W(\R^3,\C^6) \bigr ] 
\end{align*}
as well as the one-to-one correspondence between real fields and complex fields with $\omega \geq 0$ from Proposition~\ref{Schroedinger:prop:1_to_1_correspondence}, 
\begin{align}
	L^2(\R^3,\R^6) \ni (\mathbf{E},\mathbf{H}) = 2 \Re \Psi_+ 
	\; \; \longleftrightarrow \; \; 
	\Psi_+ = Q_+ (\mathbf{E},\mathbf{H}) \in \Hil_W
	. 
	\label{Schroedinger:eqn:Maxwell_one_to_one_correspondence}
\end{align}
That means Maxwell's equations~\eqref{intro:eqn:Maxwell_equations} are equivalent to the Schrödinger equation
\begin{align}
	\ii \, \partial_t \Psi_+(t) &= M_+ \Psi_+(t) - \ii \, J_+(t) 
	, 
	&&
	\Psi_+(t_0) = Q_+ \bigl ( \mathbf{E}(t_0) , \mathbf{H}(t_0) \bigr ) 
	, 
	\label{Schroedinger:eqn:Schroedinger_Maxwell_reduced}
\end{align}
where $J_+(t) := Q_+ \, W^{-1} \, \bigl ( \mathbf{J}(t) , 0 \bigr )$ is the non-negative frequency contribution to the current density and the \emph{Maxwell operator} 
\begin{align*}
	M_+ := M^{\C} \big \vert_{\Hil_W} 
\end{align*}
is the restriction of the complexified Maxwell operator to non-negative frequencies. The equivalence of the dynamical equations \eqref{Schroedinger:eqn:Schroedinger_Maxwell_reduced} and \eqref{intro:eqn:Maxwell_equations:dynamics} follows from the one-to-one correspondence~\eqref{Schroedinger:eqn:Maxwell_one_to_one_correspondence} and the equivalence of the complexified equations. Similarly, if the initial state $\bigl ( \mathbf{E}(t_0) , \mathbf{H}(t_0) \bigr )$ satisfies the constraint equations and charge is conserved, then $2 \Re \Psi_+(t)$ also satisfies the constraint equation~\eqref{intro:eqn:Maxwell_equations:constraint}. 

Lastly, the selfadjointness of $M_+$ translates to conservation of total field energy, which is the reason why in the context of electromagnetism we refer to the weighted scalar product~\eqref{Schroedinger:eqn:weighted_scalar_product} as \emph{energy} scalar product. 

\subsubsection{Extension to media with complex material weights} 
\label{Schroedinger:Maxwell_equations:extension}
For media with real weights, the (complexified) Schrödinger formalism for Maxwell's equations in media as presented above, was well-known since at least the 1960s, one of the earliest references we are aware of is due to Wilcox \cite{Wilcox:scattering_theory_classical_physics:1966} which predates \cite{Birman_Solomyak:L2_theory_Maxwell_operator:1987} by two decades. However, it seems that Maxwell's equations for media with \emph{complex} weights 
\begin{align*}
	W = \left (
	\begin{matrix}
		\eps & \chi \\
		\chi^* & \mu \\
	\end{matrix}
	\right )
\end{align*}
that satisfy Assumption~\ref{Schroedinger:assumption:material_weights} have not been derived and studied prior to \cite{DeNittis_Lein:Schroedinger_formalism_classical_waves:2017}. 

The main difficulty is to obtain physically meaningful Maxwell equations in the first place; these have to be derived from Maxwell's equations for a linear \emph{dispersive} medium \cite[Section~2]{DeNittis_Lein:Schroedinger_formalism_classical_waves:2017}. However, even if it is not at all obvious, it turns out that at the end of the day the $\omega \geq 0$ Schrödinger equation~\eqref{Schroedinger:eqn:Schroedinger_Maxwell_reduced} for media with complex material weights is exactly the same as that in the real case \cite[Section~3.3]{DeNittis_Lein:Schroedinger_formalism_classical_waves:2017}. 

%% file: section_3.tex
\section{Rigorous quantum-wave analogies in photonic crystals} 
\label{PhCs}
The Schrödinger formalism for classical waves we have summarized in the last section becomes the starting point for the actually interesting part — investigating whether and to what extent specific quantum-wave analogies hold. Examples that have been covered by other authors include Anderson localization \cite{John:photonic_crystals:1987,Figotin_Klein:localization_classical_waves_II:1997}, effective dynamics in weakly non-linear photonic crystals in the form of non-linear Schrödinger equations \cite{Babin_Figotin:nonlinear_Maxwell:2003,Babin_Figotin:nonlinear_Maxwell_4:2005}, a proof of the Bethe-Sommerfeld conjecture in two dimensions \cite{Vorobets:photonic_Bethe_Sommerfeld_conjecture:2011} as well as scattering theory for asymptotically homogeneous electromagnetic media \cite{Wilcox:theory_Bloch_waves:1978,Schulenberger_Wilcox:completeness_wave_operators:1971,Reed_Simon:scattering_theory_wave_equations:1977}. 

Our contributions to the subject  \cite{DeNittis_Lein:adiabatic_periodic_Maxwell_PsiDO:2013,DeNittis_Lein:symmetries_Maxwell:2014,DeNittis_Lein:sapt_photonic_crystals:2013,DeNittis_Lein:ray_optics_photonic_crystals:2014,DeNittis_Lein:Schroedinger_formalism_classical_waves:2017,DeNittis_Lein:symmetries_electromagnetism:2017} work towards a proof of \emph{photonic bulk-boundary correspondences} which link specific physical phenomena in photonic crystals to topology; the paradigmatic example is the one proposed by Haldane in \cite{Raghu_Haldane:quantum_Hall_effect_photonic_crystals:2008}, the seminal paper that kickstarted the search for topological effects in classical waves.

\subsection{Fundamental properties of periodic Maxwell operators} 
\label{PhCs:photonic_crystals}
The first step was to better understand the mathematical properties of \emph{photonic crystals} \cite{Kuchment:math_photonic_crystals:2001,DeNittis_Lein:adiabatic_periodic_Maxwell_PsiDO:2013}, an idea Yablonovitch \cite{Yablonovitch:photonic_crystals_genisis:1987} and John \cite{John:photonic_crystals:1987} independently came up with. Yablonovitch envisioned that periodic patterning would allow one to create a medium that acts as an “omnidirectional quarter wave plate”; in the parlance of condensed matter physics, the medium should have a \emph{photonic band gap}. What makes photonic crystals so interesting is that compared to crystalline solids physicists have much wider latitude when \emph{engineering} them for a specific purpose. Indeed, a few years after the concept was proposed, Yablonovitch successfully designed and manufactured a photonic crystal with a band gap \cite{Yablonovitch_Gmitter_Leung:photonic_band_structure_fcc_nonspherical:1991}. 
 
Mathematically speaking, photonic crystals are described by Maxwell operators with periodic weights. 
\begin{assumption}[Periodic weights]\label{PhCs:assumption:periodic_weights}
	$W$ is periodic with respect to some lattice $\Gamma \cong \Z^3$. 
\end{assumption}
Many of the standard techniques for periodic operators \cite{Kuchment:Floquet_theory:1993,Kuchment:math_photonic_crystals:2001} apply directly, and it is not surprising that periodic Maxwell oeprators admit a frequency band picture \cite[Theorem~1.4]{DeNittis_Lein:adiabatic_periodic_Maxwell_PsiDO:2013}: the usual Bloch Floquet transform decomposes periodic Maxwell operators 
\begin{align*}
	M_+ \cong \int_{\T^*}^{\oplus} \dd k \, M_+(k) 
\end{align*}
into a family of Maxwell operators acting on the Hilbert space $\Hil_W(k) \subset L^2_W(\T^3,\C^6)$ of the unit cell, that has been endowed with a weighted scalar product akin to~\eqref{Schroedinger:eqn:weighted_scalar_product}. Except for the infinitely degenerate eigenvalue $\omega = 0$ due to longitudinal gradient fields, the spectrum $\sigma \bigl ( M_+(k) \bigr ) \subseteq [0,\infty)$ is purely discrete. Just like with periodic Schrödinger operators, band functions are continuous and locally analytic away from band crossings; Bloch functions are locally analytic away from band crossings. 

Periodic Maxwell operators do possess characteristics that set them apart from periodic Schrödinger operators. In a departure from quantum theory, Maxwell operators necessarily feature \emph{two “ground state bands”} with approximately linear dispersion around $k = 0$ and $\omega = 0$. The presence of these ground state bands can be easily inferred from heuristic arguments that can be made rigorous (\cf Lemma~3.7 and Theorem~1.4 in \cite{DeNittis_Lein:adiabatic_periodic_Maxwell_PsiDO:2013}): low-frequency waves have very long wavelengths, which eventually become longer than the lattice length. Then to leading order these waves are subjected to the unit cell averaged weights $W_{\mathrm{avg}}$, and consequently, ground state frequency band and Bloch functions near $k = 0$ are well-approximated by those of the homogeneous medium with weights $W_{\mathrm{avg}}$. 

\subsection{Effective tight-binding operators for adiabatically perturbed photonic crystals} 
\label{PhCs:effective_tight_binding}
The main purpose of the technical paper \cite{DeNittis_Lein:adiabatic_periodic_Maxwell_PsiDO:2013} was to show that the Maxwell operator that models slowly (\ie \emph{adiabatically}) modulated photonic crystals can be viewed as a pseudodifferential operator, a necessary preliminary step for \cite{DeNittis_Lein:sapt_photonic_crystals:2013}. This is the \emph{photonic analog} of a work by Panati, Spohn and Teufel \cite{PST:effective_dynamics_Bloch:2003} who studied a Bloch electron subjected to slowly varying, external electromagnetic fields via \emph{space-adiabatic perturbation theory}.

\subsubsection{Original result} 
\label{PhCs:effective_tight_binding:original}
Unlike in quantum mechanics where the potentials due to the external fields are \emph{added}, perturbations naturally act \emph{multiplicatively} in photonics, so that the perturbed weights take the form 
\begin{align*}
	W_{\lambda}(x) = S(\lambda x)^{-2} \, W(x) 
	= \left (
	\begin{matrix}
		\tau_{\eps}^{-2}(\lambda x) & 0 \\
		0 & \tau_{\mu}^{-2}(\lambda x) \\
	\end{matrix}
	\right ) \, \left (
	\begin{matrix}
		\eps(x) & 0 \\
		0 & \mu(x) \\
	\end{matrix}
	\right )
\end{align*}
where $W$ are the periodic weights and $S(\lambda x)$ is a sufficiently regular perturbation (see \eg \cite[Assumption~1.2]{DeNittis_Lein:adiabatic_periodic_Maxwell_PsiDO:2013} or \cite[Assumption~1]{DeNittis_Lein:sapt_photonic_crystals:2013}). The dimensionless small parameter $\lambda \ll 1$ of this perturbation problem is ratio of the slow length scale on which the modulation varies to the lattice length. 

Then the associated Maxwell operator $M_{\lambda} = W_{\lambda}^{-1} \, \Rot$ (without frequency restriction) acting on the $\lambda$-dependent Hilbert space $\Hil_{\lambda} := L^2_{W_{\lambda}}(\R^3,\C^6)$ is a pseudodifferential operator \cite[Theorem~1.3]{DeNittis_Lein:adiabatic_periodic_Maxwell_PsiDO:2013}, and therefore \emph{space-adiabatic perturbation theory} developed by Panati, Spohn and Teufel \cite{PST:sapt:2002,PST:effective_dynamics_Bloch:2003} can be adapted. 

This perturbation scheme allowed us to approximate the full dynamics $\e^{- \ii t M_{\lambda}}$ by simpler, effective dynamics $\e^{- \ii t M_{\mathrm{eff,\lambda}}}$ for states from a given fixed frequency range of interest (see \cite[Theorem~1]{DeNittis_Lein:sapt_photonic_crystals:2013} for the precise mathematical statement), 
\begin{align}
	\e^{- \ii t M_{\lambda}} \, \Pi_{\lambda} = \e^{- \ii t M_{\mathrm{eff},\lambda}} \, \Pi_{\lambda} + \order_{\norm{\cdot}}(\lambda^{\infty})
	, 
	\label{PhCs:eqn:effective_dynamics_tight_binding}
\end{align}
where $\Pi_{\lambda} \asymp \sum_{n = 0}^{\infty} \lambda^n \, \Pi_n$ is the projection onto the almost invariant subspace associated to this spectral region; we will make this more precise below. All of these operators admit asymptotic series in $\lambda$ where each term can in principle be computed explicitly. 

Physicists commonly use the reasoning that is behind \eqref{PhCs:eqn:effective_dynamics_tight_binding} to \emph{justify effective tight-binding operators} that encapsulate the physics near interesting points in the band spectrum. 
\medskip

\noindent
The first order of business is to pick a \emph{finite frequency range of interest}. For the unperturbed, perfectly periodic photonic crystal, this is equivalent to selecting a finite family of frequency bands $\sigma_{\mathrm{rel}}(k) = \bigcup_{n \in \mathcal{I}} \bigl \{ \omega_n(k) \bigr \}$, $\mathcal{I} \subset \N$. However, for this spectral subspace to decouple, it is necessary to assume that $\sigma_{\mathrm{rel}}(k)$ is separated from all other frequency bands by a local \emph{spectral gap}; moreover, for physical as well as technical reasons, we need to exclude the ground state bands. 

Now let us turn on the perturbation. To quantify the error when comparing dynamics via \eqref{PhCs:eqn:effective_dynamics_tight_binding}, we have to choose a norm. This is not as immediate as it is in quantum mechanics, since for $\lambda \neq \lambda'$ the operators $M_{\lambda}$ and $M_{\lambda'}$ are defined on \emph{different} Hilbert spaces and the norms of $\Hil_{\lambda}$ and $\Hil_{\lambda'}$ \emph{depend on the perturbation parameter}. Instead, we represented all of the operators on a $\lambda$-\emph{in}dependent reference space, namely that on which the periodic operator lives. This way the norm we actually use for the estimates is independent of the perturbation parameter, and all norm estimates carry over to other representations even if the unitaries themselves depend on $\lambda$. 

In the end, we were able to mimic the construction of Panati, Spohn and Teufel. For example, we constructed the projection $\Pi_{\lambda} \asymp \sum_{n = 0}^{\infty} \lambda^n \, \Pi_n$ from equation~\eqref{PhCs:eqn:effective_dynamics_tight_binding} \cite[Propositions~1]{DeNittis_Lein:sapt_photonic_crystals:2013}. This projection is uniquely determined up to $\order(\lambda^{\infty})$ by three data: (1)~$\Pi_{\lambda}$ is an orthogonal projection; (2)~$\Pi_{\lambda}$ commutes with $M_{\lambda}$ up to $\order(\lambda^{\infty})$; and (3)~the leading-order term $\Pi_0 \cong \sum_{n \in \mathcal{I}} \sopro{\varphi_n(\hat{k})}{\varphi_n(\hat{k})}$ is unitarily equivalent to the projection onto the states associated to the relevant bands $\sigma_{\mathrm{rel}}(k) = \bigcup_{n \in \mathcal{I}} \bigl \{ \omega_n(k) \bigr \}$. Moreover, we verified that modulo an $\order(\lambda^{\infty})$ error, its range $\ran \Pi_{\lambda}$ is comprised of transversal waves \cite[Proposition~7]{DeNittis_Lein:sapt_photonic_crystals:2013}, so that states from $\ran \Pi_{\lambda}$ satisfy the constraint~\eqref{intro:eqn:Maxwell_equations:constraint} for $\rho = 0$. 

\subsubsection{Incorporating the restriction to $\omega \geq 0$} 
\label{PhCs:effective_tight_binding:omega_geq_0}
Strictly speaking \cite{DeNittis_Lein:sapt_photonic_crystals:2013} uses unphysical equations as it predates our more recent works \cite{DeNittis_Lein:ray_optics_photonic_crystals:2014,DeNittis_Lein:Schroedinger_formalism_classical_waves:2017} that were the first to start the analysis with the correct equations. Compared to our newer works, the Maxwell operator $M_{\lambda} = W_{\lambda}^{-1} \, \Rot$ from the older paper \cite{DeNittis_Lein:sapt_photonic_crystals:2013} lacks the restriction to non-negative frequencies. 

Nevertheless, this can be remedied with simple, straightforward arguments. First of all, instead of picking bands “symmetrically” as discussed in \cite[Section~4.1.1]{DeNittis_Lein:sapt_photonic_crystals:2013}, we only choose \emph{positive} frequency bands in the construction of $\Pi_{\lambda}$. 

To show we can replace the evolution generated by $M_{+,\lambda}$ with that of $M_{\lambda}$ on $\ran \Pi_{\lambda}$, we just combine a Duhamel argument with 
\begin{lemma}\label{PhCs:lem:M_lambda_replaced_by_M_plus_lambda}
	 In the setting of \cite[Theorem~1]{DeNittis_Lein:sapt_photonic_crystals:2013}, we have 
	\begin{align}
		M_{+,\lambda} \, \Pi_{\lambda} = M_{\lambda} \, \Pi_{\lambda} + \order_{\norm{\cdot}}(\lambda^{\infty}) 
		. 
		\label{PhCs:eqn:M_lambda_replaced_by_M_plus_lambda}
	\end{align}
\end{lemma}
This lemma can be shown by modifying the proof of \cite[Proposition~7]{DeNittis_Lein:sapt_photonic_crystals:2013}. 
\begin{proof}[Sketch]
	First we pick a small enough $\omega_0 > 0$ and smooth function $\chi : \R \longrightarrow [0,1]$ whose derivatives are compactly supported and for which 
	\begin{align*}
		\chi(\omega) \, 1_{[\omega_0,\infty)}(\omega) = 1_{[\omega_0,\infty)}(\omega) 
	\end{align*}
	holds. This gives rise to a “smoothened” version $\chi(M_{\lambda})$ of the spectral projection $1_{[\omega_0,\infty)}(M_{\lambda})$ onto frequencies $\omega \geq \omega_0 > 0$. Since $\chi$ is regular enough, $\chi(M_{\lambda})$ is in fact a \emph{pseudodifferential} operator and we obtain for $\omega_0 > 0$ small enough
	\begin{align*}
		\Pi_{\lambda} \, \chi(M_{\lambda}) = \Pi_{\lambda} + \order_{\norm{\cdot}}(\lambda^{\infty}) 
		= \chi(M_{\lambda}) \, \Pi_{\lambda} + \order_{\norm{\cdot}}(\lambda^{\infty}) 
	\end{align*}
	via Weyl calculus and the Helffer-Sjöstrand formula for operator-valued symbols.
	
	The Helffer-Sjöstrand formula on the level of operators, on the other hand, yields 
	\begin{align*}
		\chi(M_{\lambda}) &= \chi(M_{\lambda}) \; 1_{[0,\infty)}(M_{\lambda}) 
		= 1_{[0,\infty)}(M_{\lambda}) \; \chi(M_{\lambda}) 
		. 
	\end{align*}
	So starting from the right-hand side of \eqref{PhCs:eqn:M_lambda_replaced_by_M_plus_lambda}, we can insert $1_{[0,\infty)}(M_{\lambda}) \, \chi(M_{\lambda})$ at the expense of an $\order_{\norm{\cdot}}(\lambda^{\infty})$ error and obtain the claim, 
	\begin{align*}
		M_{\lambda} \, \Pi_{\lambda} &= M_{\lambda} \, 1_{[0,\infty)}(M_{\lambda}) \, \chi(M_{\lambda}) \, \Pi_{\lambda} + \order_{\norm{\cdot}}(\lambda^{\infty}) 
		\\
		&= M_{+,\lambda} \, 1_{[0,\infty)}(M_{\lambda}) \, \chi(M_{\lambda}) \, \Pi_{\lambda} + \order_{\norm{\cdot}}(\lambda^{\infty}) 
		\\
		&= M_{+,\lambda} \, \Pi_{\lambda} + \order_{\norm{\cdot}}(\lambda^{\infty}) 
		. 
	\end{align*}
\end{proof}
Combining this Lemma with \cite[Theorem~1]{DeNittis_Lein:sapt_photonic_crystals:2013} gives us a the analog of \cite[Theorem~1]{DeNittis_Lein:sapt_photonic_crystals:2013} for the physically meaningful equations. 
\begin{theorem}[Effective tight-binding dynamics]\label{PhCs:thm:effective_dynamics_new}
	Suppose the adiabatically perturbed weights are of the form in \cite[Theorem~1]{DeNittis_Lein:sapt_photonic_crystals:2013} and $\sigma_{\mathrm{rel}}(k) = \bigcup_{n \in \mathcal{I}} \bigl \{ \omega_n(k) \bigr \}$ are a finite family of bands of the unperturbed operator $M_{+,0}$ that are separated from the others by a local spectral gap (\cf \cite[Assumption~3]{DeNittis_Lein:sapt_photonic_crystals:2013}) and $0 \not\in \sigma_{\mathrm{rel}}(0)$. Moreover, we assume that the Bloch bundle (\cf \cite[Section~4.2]{DeNittis_Lein:symmetries_electromagnetism:2017}) is trivial. 
	
	Then there exists an orthogonal projection $\Pi_{\lambda} \asymp \sum_{n = 0}^{\infty} \lambda^n \, \Pi_n$ onto an almost invariant subspace associated to $\sigma_{\mathrm{rel}}$ and an effective Maxwell operator $M_{\mathrm{eff},\lambda} \asymp \sum_{n = 0}^{\infty} \lambda^n \, M_{\mathrm{eff},n}$ so that 
	\begin{align*}
		\e^{- \ii t M_{+,\lambda}} \, \Pi_{\lambda} = \e^{- \ii t M_{\mathrm{eff},\lambda}} \, \Pi_{\lambda} + \order_{\norm{\cdot}}(\lambda^{\infty})
	\end{align*}
	holds. These operators are in fact pseudodifferential operators that can be computed to any order in $\lambda$ (see \cite[equations~(8)–(10)]{DeNittis_Lein:sapt_photonic_crystals:2013}). 
\end{theorem}
Unfortunately, the price we have to pay is that our work inherits the same restriction as in \cite{PST:effective_dynamics_Bloch:2003,DeNittis_Lein:Bloch_electron:2009}: the Bloch bundle associated to $\sigma_{\mathrm{rel}}(k)$ needs to be trivial (\cite[Assumption~$\mathrm{A}_2$]{PST:effective_dynamics_Bloch:2003} or \cite[Assumption~3.2]{DeNittis_Lein:Bloch_electron:2009}). However, a more recent work of Freund and Teufel \cite{Freund_Teufel:non_trivial_Bloch_sapt:2013} shows how we might go about extending Theorem~\ref{PhCs:thm:effective_dynamics_new} to the topologically non-trivial case in the future. Their elegant approach develops a pseudodifferential calculus on vector bundles that crucially also works when the Bloch vector bundle is non-trivial. 

\subsection{Topological classification of electromagnetic media} 
\label{PhCs:topological_classification}
While our previous works were mostly focussed on Haldane's \emph{photonic bulk-edge conjecture} \cite{Raghu_Haldane:quantum_Hall_effect_photonic_crystals:2008}, a natural and for physicists perhaps more interesting question is whether there exist electromagnetic media with as-of-yet undiscovered topological phenomena. Simply put, we can say that a phenomenon is of topological origin if there is a physical observable 
\begin{align*}
	O(t) \approx T
\end{align*}
that is approximately given by a topological invariant $T$; for the Quantum Hall Effect that observable is the transverse conductivity and $T$ is the Chern number. 

Since the \emph{types} of topological invariants supported by a physical system depend on its dimensionality as well as its topological class, and the topological class is determined by 
the \emph{number and nature of its discrete symmetries}, this question can be answered by applying the standard classification tool for topological insulators, the Cartan-Altland-Zirnbauer scheme \cite{Altland_Zirnbauer:superconductors_symmetries:1997,Chiu_Teo_Schnyder_Ryu:classification_topological_insulators:2016}, to electromagnetic media.

\subsubsection{The Cartan-Altland-Zirnbauer classification of topological insulators} 
\label{PhCs:topological_classification:CAZ_scheme}
The relevant symmetries for the topological classification are unitaries or antiunitaries $V$ that \emph{square to $\pm \id$}, and \emph{either commute or anticommute} with the quantum Hamilton operator, 
\begin{align*}
	V \, H \, V^{-1} = \pm H 
	. 
\end{align*}
In the parlance of topological insulators, unitary, commuting symmetries are referred to as usual symmetries; unitary, \emph{anti}commuting symmetries are chiral symmetries; \emph{anti}unitary, commuting symmetries are time-reversal symmetries; and \emph{anti}unitary, \emph{anti}commuting symmetries are particle-hole symmetries. The two antiunitary symmetries come in an even and an odd variety depending on whether $V^2 = \pm \id$. 

Before proceeding with the classification, we need to block-diagonalize $H = \left ( 
\begin{smallmatrix}
	H_+ & 0 \\
	0 & H_- \\
\end{smallmatrix}
\right )$ with respect to unitary, commuting symmetries — if it has any — until none are left. Note that the resulting block operators $H_{\pm}$ may lose other symmetries along the way, as the other symmetries $V \neq \left ( 
\begin{smallmatrix}
	V_+ & 0 \\
	0 & V_- \\
\end{smallmatrix}
\right )$ need not be block-diagonal. For example, this is why some time-reversal symmetric electromagnetic media with a certain crystallographic symmetry have topologically protected edge modes with fixed isospin (\eg \cite{Wu_Hu:topological_photonic_crystal_with_time_reversal_symmetry:2015} and the discussion in \cite[Section~5.2.2]{DeNittis_Lein:symmetries_electromagnetism:2017}). 

\subsubsection{Material symmetries of electromagnetic media} 
\label{PhCs:topological_classification:material_symmetries}
When experimentalists design a topological photonic crystal, they have two axes to explore: (1)~they can select different materials from which to build the photonic crystal (with \emph{material symmetries}) and (2)~then decide how to periodically arrange them (\emph{crystallographic symmetries}). The primary focus of \cite{DeNittis_Lein:symmetries_Maxwell:2014,DeNittis_Lein:symmetries_electromagnetism:2017} was to obtain a classification in terms of \emph{material symmetries}. Those relate $\mathbf{E}$ and $\mathbf{H}$, and are of the form 
\begin{subequations}\label{PhCs:eqn:material_symmetries}
	\begin{align}
		U_n :& \negmedspace= \sigma_n \otimes \id 
		, 
		&& 
		n = 1 , 2 , 3 
		, 
		\label{PhCs:eqn:material_symmetries:unitary}
		\\
		T_n :& \negmedspace= (\sigma_n \otimes \id) \, C 
		, 
		&&
		n = 0 , 1 , 2 , 3 
		, 
		\label{PhCs:eqn:material_symmetries:antiunitary}
	\end{align}
\end{subequations}
where the tensor product refers to the $(\mathbf{E},\mathbf{H})$ splitting, $\sigma_0 = \id$ is the identity and $\sigma_1$, $\sigma_2$ and $\sigma_3$ are the three Pauli matrices. The usual time-reversal symmetry operator $T_3 \bigl ( \psi^E , \psi^H \bigr ) = \bigl ( \overline{\psi^E} \, , \, - \overline{\psi^H} \bigr )$, for example, complex conjugates the fields and flips the sign of the magnetic component. The form of the material symmetries is suggested by the free Maxwell operator 
\begin{align*}
	\Rot = \left (
	\begin{matrix}
		0 & + \ii \nabla^{\times} \\
		- \ii \nabla^{\times} & 0 \\
	\end{matrix}
	\right ) 
	= - \sigma_2 \otimes \nabla^{\times} 
\end{align*}
that can be written in terms of the Pauli matrix $\sigma_2$. Thus, symmetries of the form \eqref{PhCs:eqn:material_symmetries} either commute or anticommute with $\Rot$. 

Initially, we performed a symmetry analysis on the operator $M = W^{-1} \, \Rot$ that lacked the restriction to non-negative frequencies \cite{DeNittis_Lein:symmetries_Maxwell:2014}. While all of the arguments in that paper are mathematically correct, the equations we study there are unphysical. One of the mistakes we have made is that we have not taken into account that positive and negative frequency states are not independent degrees of freedom. Indeed, some of the symmetries we have considered in our earlier work, though, mixed the two — which is inadmissible. 

Put another way, symmetries $V = U_n , T_n$ of the form \eqref{PhCs:eqn:material_symmetries} need to be compatible with the restriction to $\omega \geq 0$. This admissibility condition requires that $V$ has to \emph{commute} with the Maxwell operator $M = W^{-1} \, \Rot$ that lacks the frequency restriction to $\omega \geq 0$, 
\begin{align*}
	V \, M \, V^{-1} = + M 
	, 
\end{align*}
for otherwise $V$ maps $\omega \geq 0$ states onto $\omega \leq 0$ states. Only then does $V$ restrict to a symmetry of $M_+ = M \, \vert_{\omega \geq 0}$. 

The second requirement is that $V = U_n , T_n$ needs to be also an (anti)unitary with respect to the \emph{weighted} scalar product~\eqref{Schroedinger:eqn:weighted_scalar_product}, which translates to 
\begin{align*}
	V \, W \, V^{-1} = + W
	. 
\end{align*}
Combining these two commutativity conditions with the product structure of $M_+$, we see that the only symmetries that \emph{may} play a role are those three which commute with the free Maxwell operator, namely $T_1$, $U_2$ and $T_3$ \cite[Lemma~3.1]{DeNittis_Lein:symmetries_electromagnetism:2017}. 

\subsubsection{Classification result} 
\label{PhCs:topological_classification:classification_result}
Of those three symmetries, $U_2$ is a unitary, commuting symmetry and therefore plays no role in the classification scheme. That gives us four topologically distinct electromagnetic media, each is characterized by which of the even time-reversal symmetries, $T_1$ or $T_3$, is present or broken. 
\begin{theorem}[Symmetry classification of media {{\cite[Theorem~1.4]{DeNittis_Lein:symmetries_electromagnetism:2017}}}]\label{intro:thm:classification_media}
	Suppose the material weights 
	\begin{align*}
		W(x) = \left (
		\begin{matrix}
			\eps(x) & \chi(x) \\
			\chi(x)^* & \mu(x) \\
		\end{matrix}
		\right )
		= \left (
		\begin{matrix}
			w_0(x) + w_3(x) & w_1(x) - \ii w_2(x) \\
			w_1(x) + \ii w_2(x) & w_0(x) - w_3(x) \\
		\end{matrix}
		\right )
		,
	\end{align*}
	expressed in terms of four hermitian $3 \times 3$ matrices $w_j(x) = w_j(x)^*$, $j = 0 , 1 , 2 , 3$, 
	are lossless and have strictly positive eigenvalues that are bounded away from $0$ and $\infty$, \ie they satisfy Assumption~\ref{Schroedinger:assumption:material_weights}. Moreover, we assume $M_+$ has no additional unitary commuting symmetries. Then there are \emph{four topologically distinct types of electromagnetic media}: 
	\begin{center}
		\newcolumntype{A}{>{\centering\arraybackslash\normalsize} m{2.85cm} }
		\newcolumntype{B}{>{\centering\arraybackslash\normalsize} m{2.3cm} }
		\newcolumntype{C}{>{\centering\arraybackslash\normalsize} m{2.475cm} }
		\newcolumntype{D}{>{\centering\arraybackslash\normalsize} m{1.7cm} }
		\newcolumntype{E}{>{\centering\arraybackslash\normalsize} m{1.45cm} }
		\renewcommand{\arraystretch}{1.15}
		\begin{tabular}{A | C | c | E}
			\emph{Material} & \emph{Conditions on $W$} & \emph{Symmetries} & \emph{CAZ Class} \\ \hline \hline 
			Dual symmetric materials \& vacuum & $w_0 = \Re w_0$, $w_3 = 0$, $w_1 = 0$, $w_2 = \Re w_2$ & $T_1$, $T_3$, $U_2$ & 2 $\times$ AI \\ \hline
			Non-dual symmetric \linebreak \& non-gyrotropic & $w_0 = \Re w_0$, $w_3 = \Re w_3$, $w_1 = \ii \, \Im w_1$, $w_2 = \Re w_2$ & $T_3$ & AI \\ \hline
			Magneto-electric & $w_0 = \Re w_0$, $w_3 = \ii \, \Im w_3$, $w_1 = \Re w_1$, $w_2 = \Re w_2$ & $T_1$ & AI \\ \hline
			Gyrotropic & 
			None & None & A \\ 
		\end{tabular}
	\end{center}
	The conditions on $W$ in each row are exclusive, meaning that \eg non-gyrotropic materials must violate at least one of the conditions that single out magneto-electric materials. 
\end{theorem}
For three of these types of media, the topological classification is well-known in the literature \cite{Nenciu:exponential_loc_Wannier:1983,Panati:triviality_Bloch_bundle:2006,DeNittis_Lein:exponentially_loc_Wannier:2011}. 

The case of dual-symmetric, non-gyrotropic media falls outside of standard theory and we had to perform that classification ourselves \cite[Section~4.2.4]{DeNittis_Lein:symmetries_electromagnetism:2017}. It turns out that after reducing out the unitary symmetry $U_2$, \ie considering left- and right-handed circularly polarized waves separately, only \emph{one} of the two time-reversal symmetries survives; each helicity component separately is then of class~AI (\cf \cite[Section~4.2.4]{DeNittis_Lein:symmetries_electromagnetism:2017}). 
\begin{theorem}[Topological bulk classification {{\cite[Theorem~1.5]{DeNittis_Lein:symmetries_electromagnetism:2017}}}]\label{intro:thm:bulk_classification}
	Suppose the material weights are periodic and satisfy Assumption~\ref{Schroedinger:assumption:material_weights}. 
	\begin{enumerate}[(1)]
		\item \textbf{Class A: Gyrotropic media} \\
		Phases are labelled by $\Z$-valued Chern numbers, in \\
		$d = 1$ by \emph{none} (topologically trivial), \\
		$d = 2$ by a \emph{single} first Chern number ($\Z$), \\
		$d = 3$ by \emph{three} first Chern numbers ($\Z^3$), \\
		$d = 4$ by \emph{six} first and \emph{one} second Chern number ($\Z^6 \oplus \Z$). 
		\item \textbf{Class AI: Non-dual symmetric, non-gyrotropic and magneto-electric media} \\
		In $d = 1 , 2 , 3$ these media are \emph{topologically trivial}, \ie there is a single phase. \\
		In $d = 4$, phases are labelled by a \emph{single} second Chern number ($\Z$). 
		\item \textbf{Dual-symmetric, non-gyrotropic media} \\
		In $d = 1 , 2 , 3$ these media are \emph{topologically trivial}, \ie there is a single phase. \\
		In $d = 4$, phases are labelled by \emph{two} second Chern numbers ($\Z^2$). 
	\end{enumerate}
\end{theorem}
Our classification result has two important consequences for physics: first of all, \emph{gyrotropic photonic crystals are indeed in the same topological class}, class~A, \emph{as quantum systems exhibiting the Quantum Hall Effect.} This is consistent with Haldane's conjecture that the edge modes predicted in \cite{Raghu_Haldane:quantum_Hall_effect_photonic_crystals:2008} and later observed by experimentalists \cite{Wang_et_al:edge_modes_photonic_crystal:2008} \emph{are indeed a photonic analog of the Quantum Hall Effect.} Given the wealth of experimental and theoretical evidence, though, this is not a surprising finding for physicists. 

What \emph{is} new is the insight that in $d \leq 3$ \emph{only gyrotropic materials are topologically non-trivial}, provided there are no other unitary, commuting symmetries. And \emph{the only} topological invariants supported in that situation are Chern numbers. 
In particular, because all potential time-reversal symmetries are \emph{even}, there are no electromagnetic media of class~AII and consequently, there exists \emph{no photonic analog of the Quantum \emph{Spin} Hall Effect} (\cf we refer to \cite[Section~5.2.2]{DeNittis_Lein:symmetries_electromagnetism:2017} for a detailed discussion of the literature, including works on spin-momentum locking). 

%% file: section_4.tex
\section{Conclusion and outlook} 
\label{outlook}
In summary, the Schrödinger formalism of electromagnetism and other classical waves opens the door to systematically adapting techniques from quantum mechanics to classical waves. Two specific cases were covered here, effective dynamics in adiabatically perturbed photonic crystals (Theorem~\ref{PhCs:thm:effective_dynamics_new}) and the topological classification of electromagnetic media (Theorem~\ref{intro:thm:classification_media}). This is not just relevant to mathematical physicists, but at least the latter result is new for and of immediate interest to physicists. 

Going forward, quantum-wave analogies will continue to serve as inspiration for mathematical, theoretical and experimental physicists. Experimentalists enjoy the much wider latitude with which media for classical waves can be engineered. Depending on the circumstances, they may choose the most suitable wave (acoustic, electromagnetic, etc.) and wavelength regime. For instance, the dynamics of spin wave packets can be “filmed” because their propagation speed is much lower than that of light \cite{Schneider_et_al:spin_wave_packet_dynamics_experiment:2008} — something that would be very hard to impossible to realize with a quantum system. Theoreticians can rely on quantum-wave or \emph{wave}-wave analogies to transfer insights from one physical system to another and to propose novel experiments. And mathematicians can find a whole host of interesting and non-trivial problems that are of immediate relevance to physics; I close this review by discussing a select few.

\subsection{Systematically developing mathematical techniques for Maxwell-type operators} 
\label{outlook:maxwell_type_operators}
One asset mathematicians bring to the table is the tendency to identify and exploit systematic commonalities: a whole host of wave equations (see \eg \cite[Section~6]{DeNittis_Lein:Schroedinger_formalism_classical_waves:2017}, \cite{Wilcox:scattering_theory_classical_physics:1966,Schulenberger_Wilcox:completeness_wave_operators:1971} or \cite{Nenciu_Nenciu:operators_classical_waves:2018}) can be phrased in the form of a Schrödinger-type equation, where a Maxwell-type operator (\cf Definition~\ref{other_waves:defn:Maxwell_type_operator}) takes the place of a quantum Hamiltonian. Thus, instead of treating Maxwell's equations, acoustic equations and linearized MHD equations separately, we may study all of them simultaneously by analyzing Maxwell-type operators instead. Simply put, we aim to connect properties of the weights to properties of Maxwell-type operators. When adapting methods and arguments from quantum mechanics, a few differences must be considered.

\subsubsection{Native techniques for operators with product rather than sum structure} 
\label{outlook:maxwell_type_operators:product_vs_sum}
The strategy of many earlier efforts to rigorously analyze Maxwell-type operators (\eg \cite[Theorem~20]{Kuchment_Levendorskii:spectrum_periodic_elliptic_operators:2001}) was to transform Maxwell-type operators $M = W^{-1} \, D \, \vert_{\omega \geq 0}$ (with product structure) to operators of the form $\widetilde{M} + V$ where the “potential” $V$ is connected to the commutator of the differential operator $D$ and a multiplication operator connected to the weights. To ensure that $V$ is “well-behaved” additional regularity conditions on $W$ need to be imposed (typically $\Cont^k$ with bounded derivatives up to $k$th order). However, commonly the material weights are piecewise constant functions, \ie they are only $L^{\infty}$ and not even continuous, as many media are fabricated by alternating two or more materials. Therefore, it will be necessary to develop more techniques which are “native” to operators with product structure. 

\subsubsection{The Hilbert space depends on the weights} 
\label{outlook:maxwell_type_operators:Hilbert_space}
Maxwell-type operators are naturally seen as selfadjoint operators on a Hilbert space that explicitly depends on the weights in two ways: first of all, the \emph{scalar product} and therefore the norm depends on the weights. This is important for \eg perturbation techniques where we would like to compare the evolution groups of the perturbed $M_{\lambda} = W^{-1}_{\lambda} \, D \, \vert_{\omega \geq 0}$ and unperturbed operators $M_0 = W^{-1}_0 \, D \, \vert_{\omega \geq 0}$ with one another. But because the norm on $\Hil_{\lambda}$ explicitly depends on the perturbation parameter $\lambda$, it is not immediately clear how to quantify the error in “power series expansions” in $\lambda$. Depending on the situation, conceptually different choices are sensible (compare \eg the approaches in \cite[Section~2.2]{DeNittis_Lein:adiabatic_periodic_Maxwell_PsiDO:2013} and \cite[Section~4]{Schulenberger_Wilcox:completeness_wave_operators:1971}). Secondly, it is not at all clear whether \eg $\e^{- \ii t M_0}$ is even well-defined as an operator on $\Hil_{\lambda}$ in the first place: \emph{a priori} we do not know \emph{whether it maps non-negative frequency states of $M_{\lambda}$ onto non-negative frequency states.} This consideration was crucial in identifying how to represent material symmetries on the complex Hilbert space in Section~\ref{PhCs:topological_classification:material_symmetries}. 

\subsubsection{Finding physically meaningful mathematical problems and interpreting mathematical results physically} 
\label{outlook:maxwell_type_operators:physics_cap_math}
A third consideration concerns conceptual differences in the physical interpretations. Mathematical physicists working on quantum systems have internalized a lot of concepts that do not always transfer as-is to classical waves. This concerns both directions, namely when giving physical meaning to mathematical statements and when translating physics to concrete mathematical problems. 

Mathematically, it is completely legitimate to study the operator $M_{\Phi} = W^{-1} \, D + \sum_{j = 1}^3 \Phi_j \, x_j$ in analogy to the Stark operator \cite{Nenciu:effective_dynamics_Bloch:1991}, and apply techniques from linear response theory \cite{DeNittis_Lein:linear_response_theory:2017}. However, adding a “potential” makes no physical sense in the context of Maxwell's equations. Instead, in experiment there are two ways to “drive the transmission” of electromagnetic fields: we can either insert sources such as an antenna \cite{Babin_Figotin:nonlinear_Maxwell_4:2005}; or we can modulate the weights (as was the case in \cite{Raghu_Haldane:quantum_Hall_effect_photonic_crystals:2008,DeNittis_Lein:sapt_photonic_crystals:2013}). Depending on the precise physical setting, either approach may make sense. 

Also in the context of topological “insulators” a tailor-made interpretation of the results is important. For quantum mechanical models for systems that exhibit the Integer Quantum Hall Effect, the Fermi projection $P_{\mathrm{F}} = 1_{(-\infty,E_{\mathrm{F}}]}(H)$ is to be interpreted as the state of the system at zero temperature. Therefore, the bulk-boundary correspondence that furnishes an explanation for the Quantum Hall Effect \cite{Thouless_Kohmoto_Nightingale_Den_Nijs:quantized_hall_conductance:1982,Hatsugai:Chern_number_edge_states:1993,Hatsugai:edge_states_Riemann_surface:1993,Prodan_Schulz_Baldes:complex_topological_insulators:2016} is a statement about how the states at the boundary relate to states in the bulk. However, in electromagnetism the “Fermi projection” $P_{\mathrm{F}} = 1_{(-\infty,\omega_{\mathrm{F}}]}(M)$, a perfectly well-defined operator, is not linked to a state and enters Haldane's photonic bulk-boundary correspondence merely as an \emph{auxiliary quantity}. 

\subsection{Dispersive media} 
\label{outlook:dispersion}
Another important feature that distinguishes classical waves from quantum systems is dispersion. To be precise here, we are not referring to the characteristic broadening of wave packets under the time evolution that also occurs in quantum system. Instead, the constitutive relations that express the auxiliary fields $(\mathbf{D},\mathbf{B})$ at time $t$ in terms of the physical fields $(\mathbf{E},\mathbf{H})$ not only depend on the instantaneous field configuration at time $t$, but also on the field configuration in the past — the medium has a “memory” (see \cite[equation~(2.3)]{DeNittis_Lein:Schroedinger_formalism_classical_waves:2017} for an explicit formula). This type of dispersion implies that initial values to Maxwell's equations for dispersive media are then \emph{past trajectories} in contrast to the setting covered here where only the instantaneous field configuration at $t = t_0$ matters. 

Within the context of topological insulators, Silveirinha has proposed to use dispersion in homogeneous instead of periodic patterning to \emph{open spectral gaps}, and explores in a systematic fashion \cite{Silverinha:Z2_topological_insulator_continuous_photonic_materials:2016,Gangaraj_Silveirinha_Hanson:Berry_connection_Maxwell:2017,Silveirinha:topological_bulk_classification_dispersive_media:2018} whether topologically non-trivial bulk band gaps exist. While there is no experimental evidence at this point and the mathematics is not well-understood, this is certainly an intriguing idea \emph{with no quantum analog} that merits further study. 

\subsection{Non-linear media} 
\label{outlook:non_linear}
The fundamental equation of quantum mechanics, the Schrödinger equation, is manifestly linear; non-linear equations like Gross-Pitaevskii \cite{Spohn:kinetic_equations_Hamiltonian_dynamics:1980,Spohn:large_scale_dynamics_interacting_particles:1991} or Hartree-Fock equations \cite[Chapter~IV.5]{Grosso_Parravicini:solid_state_physics:2003} often emerge as \emph{effective} single particle equations from linear many-body quantum mechanics. On the other hand many media of classical waves are \emph{manifestly non-linear} and it is the \emph{linear} equations that are approximations. 

That raises the question whether certain non-linear media allow us to gain insights into many-body quantum systems. For example, Babin and Figotin have shown that in a particular scaling the non-linear Maxwell equations can be approximated by a \emph{non-linear Schrödinger equation} \cite{Babin_Figotin:nonlinear_Maxwell_4:2005}, in essence allowing physicists to engineer specific non-linearities. 

Babin and Figotin's works \cite{Babin_Figotin:nonlinear_Maxwell:2003,Babin_Figotin:nonlinear_Maxwell_1:2001,Babin_Figotin:nonlinear_Maxwell_2:2002,Babin_Figotin:nonlinear_Maxwell_3:2003,Babin_Figotin:nonlinear_Maxwell_4:2005} precede that of Raghu and Haldane \cite{Raghu_Haldane:quantum_Hall_effect_photonic_crystals:2008}, so topological phenomena were not on their radar at the time. Hence, the question whether novel \emph{non-linear} topological phenomena exist is also still open. 